\documentclass[onecolumn, useAMS,usenatbib]{mn2e}
\usepackage{txfonts}
\usepackage{natbib}
\usepackage[all]{xy}
\usepackage[dvips]{graphicx}

\def\del#1{{}}
\def\gerot#1{{ #1}}

\newcommand{\ltsima}{$\; \buildrel < \over \sim \;$}
\newcommand{\lsim}{\lower.5ex\hbox{\ltsima}}
\newcommand{\gtsima}{$\; \buildrel > \over \sim \;$}
\newcommand{\gsim}{\lower.5ex\hbox{\gtsima}}
\newcommand{\bra}{\langle}
\newcommand{\ket}{\rangle}

\newcommand{\e}{\mathrm{e}}
\newcommand{\p}{\mathrm{p}}

\newcommand{\dirac}{\delta_\mathrm{D}}

\newcommand{\be}{\begin{equation}}
\newcommand{\ee}{\end{equation}}
\newcommand{\ba}{\begin{eqnarray}}
\newcommand{\ea}{\end{eqnarray}}
\newcommand{\nn}{\nonumber}
\newcommand{\mb}{\mathbf}
\newcommand{\ph}{\varphi} 
\newcommand{\bk}{\mb k}
\newcommand{\bq}{\mb q}
\newcommand{\bp}{\mb p}
\newcommand{\br}{\mb r}
\title[Perturbation Theory Trispectrum in the Time Renormalisation Approach]
{Perturbation Theory Trispectrum in the Time Renormalisation Approach}
\author[Gero J{\"u}rgens and Matthias Bartelmann]{
Gero J{\"u}rgens and Matthias Bartelmann\\
Institut f{\"u}r theoretische Astrophysik, Zentrum f{\"u}r Astronomie, Universit{\"a}t Heidelberg, Albert-Ueberle-Stra{\ss}e 2, 69120 Heidelberg, Germany}
\begin{document}
\pagerange{\pageref{firstpage}--\pageref{lastpage}}
\pubyear{2012}
\maketitle
\label{firstpage}

\begin{abstract}
An accurate theoretical description of structure formation at least in the mildly non-linear regime is essential for comparison with data from next generation galaxy surveys. 
In a recent approach one follows the time evolution of correlators directly and finds a hierarchy of evolution equations with increasing order \citep{Pietroni2008}.
So far, in this so called time renormalisation group method the trispectrum was neglected in order to obtain a closed set of equations. In this work we study the 
influence of the trispectrum on the evolution of the power spectrum. In order to keep the numerical cost at a manageable level we use the tree-level 
trispectrum from Eulerian perturbation theory. In comparison to numerical simulations we find improvement in the mildly non-linear 
regime up to $k\simeq 0.25 \, h\mathrm{Mpc}^{-1}$. Beyond $k\simeq 0.25 \, h\mathrm{Mpc}^{-1}$ the perturbative description of the 
trispectrum fails and the method performs worse than without the trispectrum included. 
Our results reinforce the conceptual advantage of the time renormalisation group method with respect to perturbation theory.
\end{abstract}
%
%

\section{Introduction}\label{sect_intro}
In the contemporary picture of our Universe structures evolve from nearly Gaussian distributed small perturbations in the homogeneous 
density field. Sound waves formed in the coupled photon-baryon fluid before recombination left oscillatory features in the matter power 
spectrum, so called baryonic acoustic oscillations (BAO). Detections of this effect have become a valuable tool to constrain cosmological 
parameters since amplitude and position of the oscillations depend on the expansion history of the Universe
\citep{Eisenstein2005,Hutsi2006,Eisenstein2007,Padmanabhan2007, Blake2007}. 
Current and upcoming galaxy surveys - such as BOSS\footnote{http://www.sdss3.org/surveys/boss.php},WFIRST
\footnote{http://wfirst.gsfc.nasa.gov}, HETDEX \citep{Hill2008} and WFMOS \citep{Glazebrook2005} - 
will measure the power spectrum of the matter distribution to 1\% accuracy in the region of the 
baryonic oscillations $0.05 \,h\mathrm{Mpc}^{-1} < k < 0.25 \,h\mathrm{Mpc}^{-1}$ \citep{Eisenstein1998,Seo2003}.

Since mode-coupling effects can significantly influence the position of the first peak at low redshifts \citep{Crocce2008}, it is necessary to find a robust 
theoretical description of structure formation in the linear and mildly non-linear regime.
While standard perturbation theory \citep{Bernardeau2002} is a powerful tool for comparison with observations from galaxy surveys on large scales 
\citep{Jeong2006,Jeong2009}, it breaks down at the scales of baryonic acoustic oscillations \citep{Jain1994}.
 
The increase of computational power and efficiency of algorithms made $N$-body simulations the most established approach for structure formation 
\citep{Springel2005,Huff2007, Heitmann2008, Takahashi2008, Evrard2008, Heitmann2010}. However, to extract statistical information from numerical simulations, either large 
sets of initial conditions or large volumes are needed and it is difficult to control measurement uncertainties in the mildly non-linear regime \citep{Angulo2008}. 
In a recent work a simple physically motivated picture was used to reduce the sample variance to speed up the scanning for cosmological parameters with 
$N$-body simulations \citep{Tassev2011}, but it remains numerically expensive and we favour a fast semi-analytical tool to predict non-linear structure formation. 

While the halo model approach \citep{Peacock1996,Smith2003} was found to be uncapable of reaching the required accuracy \citep{Huff2007}, different attempts to 
include corrections of specific types to all orders at the same time have been presented over the last years. 
Field theoretical techniques motivated by the renormalisation group \citep{Matarrese2007,Matarrese2008, Anselmi2011} and resummation methods 
\citep{Crocce2006,Crocce2006a} improved the results for power spectra in the mildly non-linear regime 
significantly down to $z=0$ in comparison to $N$-body simulations. In the following also higher order statistics were studied in these frameworks 
\citep{Valageas2008, Bernardeau2008, Guo2009}. However, these approaches are formulated for an Einstein-de~Sitter cosmology and are later generalised to other 
cosmologies by substituting the respective growth function. The accuracy of this approximation is hard to quantify at higher orders \citep{Bernardeau2002}. 

Correlating the structure formation equations with fields to different orders leads to an infinite hierarchy of evolution equations for correlators with 
increasing order. This hierarchy - similar to the well known BBGKY hierarchy \citep{Peebles1980} - was truncated at the level of the trispectrum to obtain 
a closed set of equations \citep{Pietroni2008}. This so called time renormalisation group method (TRG) can easily be generalised to a large set of different 
cosmologies including models with scale dependent 
growth functions. For example, this is important in the case of massive neutrinos \citep{Lesgourgues2006} which was also studied within this framework 
\citep{Lesgourgues2009}.

The objective of this work is to study the effect of the trispectrum on the TRG approach. Including the entire time evolution of the trispectrum would lead to an 
immoderate numerical effort. We therefore include the tree-level trispectrum from perturbation theory into the evolution equation of the 
bispectrum and study its effects on the power spectrum in the BAO regime. We work in a standard $\Lambda$CDM cosmology close to the best-fitting cosmology 
($\Omega_\mathrm M =0.25$, $\Omega_\mathrm b h^2 = 0.0224$, $h = 0.72$, $n = 0.97$ and $\sigma_8 = 0.8$) and compare to power spectra obtained from 
$N$-body simulations of the same cosmology \citep{Carlson2009}.

The paper is organised as follows: In Section~\ref{sect_trg} we review the time renormalisation approach introduced by \cite{Pietroni2008}. How the perturbation 
theory trispectrum can be included into this method will be subject of Section~\ref{sect_trispectrum}. In Section~\ref{sect_diagrams} we include the trispectrum 
into the formal analytic solution of the system and discuss the additional corrections in a diagrammatic representation and in 
Section~\ref{sect_results} our numerical results and their comparisons to $N$-body simulations are presented. 
The results are summarised and discussed in Section~\ref{sect_summary}.

\section{Time Renormalisation}\label{sect_trg}
In this section we write the structure formation equations in 
a compact matrix form and review the time renormalisation approach \citep{Pietroni2008} as a starting point for our further calculations. 
We will concentrate on spatially flat cosmologies with a dark matter component and a non-clustering dark energy fluid or $\Lambda$CDM. However, 
this method can be easily extended to more exotic cosmologies including cosmologies with a scale-dependent growth function $D_+(k,a)$ \citep{Pietroni2008}.

\subsection{Non-linear Structure Formation}\label{sect_nsf}
The time evolution of the dark matter density contrast $\delta(\mb x, \tau)$, the peculiar velocity $\mb v(\mb x, \tau)$ and 
the fluctuation of the gravitational potential $\Phi(\mb x, \tau)$ is governed by the continuity-, Euler- and Poisson equations:
\ba
\partial_\tau \, \delta(\mb x) + \nabla \cdot [(1+\delta(\mb x))\,\,\mb v(\mb x)] &=&0\\
\partial_\tau \,\mb v(\mb x) + \mathcal{H}\, \mb v(\mb x)  + \left(\mb v(\mb x) \cdot \nabla\right)\,\mb v(\mb x)  &=& -\nabla \,\Phi (\mb x)\\
\nabla^2 \Phi(\mb x) &= &\frac 32 \,\mathcal{H}^2 \, \Omega_\mathrm m \, \delta (\mb x)\hspace{0.1 cm}.
\ea
Here, the Hubble function $\mathcal H$ is the logarithmic derivative of the scale factor with respect to conformal time: $\mathcal H = \mathrm d 
\ln a / \mathrm d \tau$. For notational clarity the time dependences will be omitted for most of the quantities throughout the whole work. 
In the case of a one-component dark matter fluid we have $\Omega_\mathrm m= 1$, while in presence of an 
additional dark energy fluid we only have to replace $\Omega_\mathrm m$ by
\be
\Omega_\mathrm m = \left(1+\frac{\rho^0_\mathrm{de}}{\rho^0_\mathrm{m}}\, a^{-3\, w}\right)^{-1}\hspace{0.1 cm}.
\ee
The quantities $\rho^0_\mathrm{de}$ and $\rho^0_\mathrm{m}$ represent the background densities of the dark energy and dark matter fluid components respectively. 
$w$ denotes the equation of state parameter of the dark energy fluid, $\p_\mathrm{\,de}= w\,\rho^0_\mathrm{de}$.
Models with a cosmological constant, as for example in $\Lambda$CDM, can then simply be described by $w=-1$. 
We assume a irrotational peculiar velocity field, $\nabla \times \mb v(\mb x)=0$. Due to Helmholtz's theorem 
the velocity field can then be described by its divergence only, $\theta(\mb x)= 
\nabla \cdot \mb v(\mb x)$. For the two remaining scalar fields $\theta(\mb x)$ and $\delta(\mb x)$ the evolution in Fourier space 
reads \citep{Bernardeau2002}:
\ba
\label{fourierevo}
\partial_\tau \,\delta(\mb k) + \theta(\mb k) &=& - \dirac(\mb k- \mb q -\mb p)\,\,\alpha (\mb q, \mb p)\, \,\theta(\mb q)\,\, \delta (\mb p)  \nn \\
\partial_\tau \,\theta(\mb k) + \mathcal H \,\theta(\mb k) + \frac 32 \,\mathcal H ^2\, \Omega_\mathrm m \, \delta(\mb k)  &=&
- \dirac(\mb k- \mb q -\mb p)\,\,\beta (\mb q, \mb p)\,\, \theta(\mb q)\,\, \theta (\mb p)\hspace{0.1 cm}.
\ea
Here and in the following, repeated momentum arguments in products (for instance $\mb p,\mb q$ in eqn.~(\ref{fourierevo})) 
imply an integration over the respective momenta $ \int \mathrm d^3 \mb p \int \mathrm d^3 \mb q...$ . The left hand side of eqn.~(\ref{fourierevo}) 
represents the linear evolution of the fields while the right hand side 
expresses the non-linear mode-coupling, which is determined by the two model independent mode-coupling functions
\ba
\alpha(\mb q, \mb p)&=& \frac{(\mb p+\mb q)\cdot\mb q}{q^2} \hspace{0.1 cm},\nn \\
\beta(\mb q, \mb p) & = & \frac{(\mb p+\mb q)^2\,\mb p\cdot\mb q}{2\,q^2 p^2}\hspace{0.1 cm}.
\ea 
To write the equations in a more compact form, we introduce the logarithmic time variable $\eta$, which explicitly 
contains the linear structure growth with respect to the scale factor:  
\be
\eta = \ln \, \left(\frac{D_+(a)}{D_+(a_\mathrm{in})}\right)\hspace{0.1 cm}.
\ee
The scale factor $a_\mathrm{in}$ may be chosen at a time when the systen could still be well aproximated to be Gaussian and in the linear regime. 
Now a doublet field can be introduced for the density contrast $\delta(\mb k)$ and the divergence of the velocity dispersion $\theta(\mb k)$:
\be
\left(\begin{tabular}{r}$\ph_1(\mb k)$\\
		    $\ph_2(\mb k)$
		      \end{tabular}\right)= \e^{-\eta}\,\, \left(\begin{tabular}{c}$\delta\,(\mb k)$\\
		    $-\theta\,(\mb k)/(\mathcal H \, f_+)$
		      \end{tabular}\right)  \hspace{0.1 cm}.
\ee
Here, $f_+$ is the logarithmic derivative of the growth function $D_+$ with respect to the scale factor $a$, $f_+ = \mathrm d \ln D_+/ \mathrm d \ln a$. 
The factor $\e^{-\eta}$ now compensates for the linear evolution of the fields $\ph_i(\mb k)$. In 
other words, solving the linearised structure formation equations for these fields would lead to no time dependence in the doublet field at all. 
Therefore, any evolution away from 
the initial field douplet is explicitly due to non-linear effects. 
The set of equations~(\ref{fourierevo}) can now be expressed in a very compact form:
\be
\label{sfe_compact}
\partial_\eta\, \ph_a(\mb k) = \Omega _{ab}\, \ph_b(\mb k) + \e^\eta \,\,\tilde \gamma_{abc}\,(\mb k,-\mb q,-\mb p)\, \ph_b(\mb q)\, \ph_c(\mb p)\hspace{0.1 cm}.
\ee
Here, the linear evolution is governed by the matrix
\be
 \Omega_{ab} = \left(\begin{tabular}{c | c}$1$&$-1$ \\  $-3\,\Omega_\mathrm m / (2\, f_+^2)$&$3\,\Omega_\mathrm m / (2\, f_+^2)$
     \end{tabular}\right) \hspace{0.1 cm},
\ee
while the non-linear mode-coupling is moderated by the vertex funtions $\tilde \gamma_{abc}\,(\mb k, \mb q, \mb p)$. The only non-vanishing vertex functions are 
\ba
\label{vertex}
\tilde\gamma_{121}\,(\mb k, \mb q, \mb p) = \tilde\gamma_{112}\,(\mb k, \mb p, \mb q) &=& \dirac (\mb k+ \mb q+\mb p)\,\,\frac{ \alpha\,(\mb q, \mb p)}2\hspace{0.1 cm },\nn \\
\tilde\gamma_{222}\,(\mb k, \mb q, \mb p) &=& \dirac (\mb k+ \mb q+\mb p)\,\,\beta\,(\mb q, \mb p)\hspace{0.1 cm }.
\ea
Since the vertex functions only appear in integrals, it is notationally convenient to introduce vertex functions, for which the $\dirac$-function is 
already integrated out: 
\be
\gamma_{acb}^{\,k,q,|\mb k + \mb q|} \equiv \int \mathrm d^3 p \,\, \tilde \gamma_{abc}(\mb k, \mb q, \mb p)\hspace{0.1 cm}.
\ee
These quantities turn out to dependent only on the absolute values of the wave vectors $k,q$ and $|\bk+\bq|$. 
\subsection{Hierarchy of Correlators}\label{sect_hierarchy}
While in standard perturbation theory one aims to solve the evolution equations of the fields themselves, in the time renormalisation approach one 
formulates evolution equations directly for the final quantities of interest - the correlators of the fields.
In order to do this, one can use eqn.~(\ref{sfe_compact}) to write down a hierarchy of evolution equations for correlators of any order:
\ba
\label{hierarchy}
\partial_\eta \bra \ph_a \ph_b \ket &=& -\Omega_{ac}\bra \ph_c \ph_b \ket - \Omega_{bc}\bra \ph_a \ph_c \ket 
                                      + \e^\eta  \left[ \tilde \gamma_{acd}\bra \ph_c \ph_d \ph_b \ket + \tilde \gamma_{bcd}\bra \ph_a \ph_c \ph_d \ket \right] \nn \\
\partial_\eta \bra \ph_a \ph_b \ph_c \ket &=& -\Omega_{ad}\bra \ph_d \ph_b \ph_c \ket + \mathrm{c.p.}\{a,b,c\}
                                      + \e^\eta \left[ \tilde \gamma_{ade}\bra  \ph_d \ph_e \ph_b \ph_c\ket + \mathrm{c.p.}\{a,b,c\}\right]\nn \\
\partial_\eta \bra \ph_a \ph_b \ph_c \ph_d \ket &=&  ...\hspace{0.1 cm}.
\ea
Here and in the following, we abbreviate the doublet field index $a_1$ and the wave vector $\bk_1$ to a single number index, i.e. $\ph_1 \equiv \ph_{a_1} (\bk_1)$.
As a natural property of this hierarchy, for the evolution of a correlator of order $n$ the knowledge of correlators of the next higher order $n+1$ is needed. 
Therefore, one is obliged to truncate this hierarchy at a certain point in order to obtain a closed set of equations. 
Splitting up the four-point correlator into its connected and unconnected part yields by Wick's theorem
\be
\label{four-point}
\bra \ph_1 \ph_2 \ph_3 \ph_4\ket =  \bra \ph_1 \ph_2 \ph_3 \ph_4\ket_\mathrm c
 + \bra \ph_1\ph_2\ket \bra \ph_3\ph_4\ket + \bra \ph_1\ph_3\ket \bra \ph_2\ph_4\ket + \bra \ph_1\ph_4\ket \bra \ph_2\ph_3\ket\hspace{0.1 cm}. 
\ee
For instance, one can close the system by neglecting the connected part of the four-point correlator, as it was done by \cite{Pietroni2008}. 
Due to this approximation, one is left with the first two equations of the hierarchy and the simplified system is then fully described by its
power spectra $P_{ab}^{\,k}$ and its bispectra $B_{abc}^{\,k_1,k_2,k_3}$
\ba
\bra \ph_1 \ph_2\ket &=& \dirac\,(\mb k_1 + \mb k_2)\, P_{ab}^{\,k_1}\nn \\
\bra \ph_1 \ph_2 \ph_3\ket &=& \dirac\,(\mb k_1 + \mb k_2+ \mb k_3) \,B_{abc}^{\,k_1, k_2,k_3}\hspace{0.1 cm}.
\ea
Due to isotropy the bispectrum will only depend on the absolute values of the wave vectors.
Integrating eqn.~(\ref{hierarchy}) over one wave vector and using eqn.~(\ref{four-point}) with $\bra \ph_1 \ph_2 \ph_3 \ph_4\ket_\mathrm c=0$ one finds a 
closed system of equations in which the four-point function is represented in terms of power spectra $P_{ab}^{\,k}$:
\ba
\label{system}
\partial_\eta P_{ab}^{\,k} &=& -~\Omega_{ac}\,P_{cb}^{\,k} -\Omega_{bc}\,P_{ac}^{\,k} + \e^\eta  \int \mathrm d^3 q 
\left[\gamma_{acd}^{\,k,q,p} B_{bcd}^{\,k,q,p} +  (a\leftrightarrow b)\right]\nn \\
\partial_\eta B_{abc}^{\,k,q,p}&=&-~\Omega_{ad}B_{dbc}^{\,k,q,p}-\Omega_{bd}B_{adc}^{\,k,q,p} -\Omega_{cd}B_{abd}^{\,k,q,p} + ~2\, \e^\eta\,
\left[\gamma_{ade}^{\,k,q,p}P_{db}^{\,q}P_{ec}^{\,p}+\gamma_{bde}^{\,q,p,k}P_{dc}^{\,p}P_{ea}^{\,k}+\gamma_{cde}^{\,p,k,q}P_{da}^{\,k}P_{eb}^{\,q}\right]\hspace{0.1 cm}.
\ea
Here and in the following $\bp$ will denote the vector $\bp=-(\bk+\bq)$. How this system can be solved numerically will be subject of the next subsection.
\subsection{Solving the Closed System}\label{sect_solving}
Formal and numerical solutions to the closed system in eqn.~(\ref{system}) have already been presented \citep{Pietroni2008}. We follow the same numerical path 
to investigate the solution's sensitivity to the perturbation theory trispectrum. Since we are mainly interested in the evolution of the power spectrum 
itself, one circumvents the necessity of tracking the total bispectrum by introducing auxiliary integrals
\be
I_{acd,bef}^{\,k} \equiv \frac{k}{4 \,\pi} \int \mathrm{d}^3 q \, \frac 12 \left[ 
\gamma_{acd}^{\,k,q,p}\, B_{bef}^{\,k,q,p} + (q\leftrightarrow p)\right]\hspace{0.1 cm}.
\ee
The introduction of these integrals encapsulates the one loop character of the power spectrum evolution equation (first equation in (\ref{system})), which simplifies to the tree level 
equation
\be
\label{p_evolution}
\partial_\eta P_{ab}^{\,k}= -\Omega_{ac}P_{cb}^{\,k}-\Omega_{bc}P_{ab}^{\,k}
+\e^\eta \frac{4\, \pi}{k} \, \left[I_{acd,bcd}^{\,k}+I_{bcd,acd}^{\,k}\right]\hspace{0.1 cm}.
\ee
Differentiating the integrals $I_{acd,bef}^{\,k}$ with respect to time $\eta$ and using the evolution equation of the bispectrum from eqn.~(\ref{system}) gives the 
following time evolution of these integrals:
\be
\label{i_evolution}
\partial_\eta I_{acd,bef}^{\,k} = -\Omega_{bg}I_{acd,gef}^{\,k}-\Omega_{eg}I_{acd,bgf}^{\,k}-\Omega_{fg}I_{acd,beg}^{\,k} + 2\, \e^\eta\, A_{acd,bef}^{\,k}\hspace{0.1 cm}.
\ee
The $k$-space loop integral which was originally in the evolution of the power spectrum now appears in the mode-coupling integrals $A_{acd,bef}^{\,k}$, 
which initially drive the system away from Gaussianity:
\be
\label{a_integral}
A_{acd,bef}^{\,k} =  \frac{k}{4 \,\pi} \int \mathrm{d}^3 q \frac 12 \left\{ \gamma_{acd}^{\,k,q,p}  \left(
\gamma_{bgh}^{\,k,q,p}P_{ge}^{\,q}P_{hf}^{\,p}  +\gamma_{egh}^{\,q,p,k}P_{gf}^{\,p}P_{hb}^{\,k}+\gamma_{fgh}^{\,p,k,q}P_{gb}^{\,k}P_{hb}^{\,q}\nn\\
  \right) + (q \leftrightarrow p)\right\}\hspace{0.1cm}.
\ee
The calculation of the integrals $A_{acd,bef}^{\,k}$ is so far the only time consuming task in numerically solving the system. 
\subsection{Initial Conditions and Symmetries}\label{sect_symmetries}
We propagate the system of equations~(\ref{p_evolution})-(\ref{i_evolution}) forward in time starting from an initial time at which the dynamics of the fields 
could still be well approximated by the linearised evolution equations. As initial redshift we choose $z_\mathrm{in}=100$ and start with the linear power spectrum 
as initial contitions:
\be
\label{plin}
 P_{ab}^{\,k}(\eta=0) = u_a\, u_b\, P_\mathrm L ^{\,k}(\eta=0)\hspace{0.1 cm}, \hspace{1cm} (u_1,u_2) = (1,1)\hspace{0.1 cm}.
\ee 
Furthermore, we will assume Gaussian initial conditions. This implies a vanishing inititial bispectrum $B_{abc}^{\,k,q,p}$ and therefore also 
vanishing initial integrals $I_{acd,bef}^{\,k}$:
\be
 I_{acd,bef}^{\,k}(\eta=0) =  B_{abc}^{\,k,q,p}(\eta=0) =0\hspace{0.1 cm}.
\ee
Following the next argments, the full system of 64 integrals $I_{acd,bef}^{\,k}$ and 3 power spectra $P_{ab}^{\,k}$ can be reduced to 14 independent components.
Since the only non-vanishing vertex contributions appear for the index triples $(acd)\in\{(112),(121),(222)\}$ only the integrand $I_{acd,bef}^{\,k}$ with 
these triples will evolve away from zero, which can be seen from equations~(\ref{i_evolution})-(\ref{a_integral}). The remaining 24 components 
can be further reduced by symmetry arguments. Using the following symmetries in the vertex functions and the bispectrum
\be
\gamma_{acd}^{\,k,q,p}=\gamma_{adc}^{\,k,p,q}\hspace{0.1 cm}, \hspace{1cm}B_{bef}^{\,k,q,p}=B_{bfe}^{\,k,p,q}\hspace{0.1 cm},
\ee
we can find the following symmetry for the integrals $I_{acd,bef}^{\,k}$:
\be
I_{acd,bef}^{\,k} = I_{adc,bfe}^{\,k}\hspace{0.1 cm}.
\ee
Due to this symmetry, only 14 independent integrals remain to be followed. The independent integrals are identified by the direct product of 
$(acd)=(112)$ and $(bef)= (b11)$,$(b12)$, $(b21)$ and $(b22)$, $(b=1,2)$, and the direct product of 
$(acd)=(222)$ and $(bef)= (b11)$,$(b12)$ and $(b22)$, $(b=1,2)$. Including the 3 independent power spectra $P_{11}^{\,k}$, $P_{12}^{\,k}$ and $P_{22}^{\,k}$ 
implies a system of 17 components in total.

\section{Trispectrum}\label{sect_trispectrum}
The connected part of the four-point correlator is called the trispectrum. To investigate the method's sensitivity with respect to the trispectrum, we 
include the perturbation theory trispectrum to third order in the linear power spectrum $P_\mathrm L^{\,k}$.
Including the full time evolution of the non-perturbative trispectrum would increase the numerical effort disproportionally. This would be 
given by the third equation in the hierarchy of time evolution equations for correlators in eqn.~(\ref{hierarchy}). However, the time evolution of the
tree-level perturbation theory trispectrum is exclusively given in terms of the growth function $D_+(a)/D_+(a_\mathrm{in})= \e^\eta$. Therefore, using this 
approximation for the bispectrum it is sufficient to compute trispectrum corrections at the initial time $a_\mathrm{in}$. These corrections can then be included 
into the routine without increasing the computational cost significantly.

\subsection{Perturbation Theory}\label{sect_pt}
In the following we adapt standard perturbation theory \citep{Bernardeau2002} to the compact matrix formulation of structure formation. 
We expand the fields $\ph_a(\bk, \eta)$ in $n$th-order perturbative contributions $\ph_a^{(n)}(\bk, \eta)$, 
which can be written in terms of the perturbation theory 
kernels $F^{(n)}_a(\bk_1,...,\bk_n)$ and the initial linear fields $\ph^{(1)}_a(\bk) = \ph^{\,\mathrm L}_a(\bk, \eta=0)$:
\ba
 \ph_a (\bk, \eta) &=& \sum_{n=1}^{\infty} \, \e^{(n-1)\,\eta} \,\ph_a^{(n)}(\bk) \label{pt-expansion}\\
 \ph_a^{(n)}(\bk) &=& \int\mathrm d^3 q_1\, ...\int\mathrm d^3 q_n\, \dirac(\bk-\bq_{1...\,n})\,F^{(n)}_a(\bq_1,...,\bq_n)\, 
\ph^{(1)}_a(\bq_1)\,...\,\ph^{(1)}_a(\bq_n)\label{pt-modes}
\ea
with $\mb q_{1...\,n}=\mb q_1+ ... +\mb q_{\,n}$. For notational simplicity we combined the two standard kernels of each order into a vector, since we will also 
need trispectrum correlations to the velocity field components:
\be
 F^{(n)}_a(\bk_1,...,\bk_n)=  \left(\begin{tabular}{r}$F^{(n)}(\bk_1,...,\bk_n)$\\
		    $G^{(n)}(\bk_1,...,\bk_n)$
		      \end{tabular}\right)  \hspace{0.1 cm}.
\ee
By inserting eqs.~(\ref{pt-expansion}-\ref{pt-modes}) into the evolution equations~(\ref{sfe_compact}) one finds the following recursion relations 
by combinatorics \citep{Goroff1986,Jain1994}:
\ba
  F^{(n)}(\bk_1, ... , \bk_n) &=& \sum_{m=1}^{n-1}\frac{G^{(m)}(\bk_1, ... , \bk_m)}{(2n+1)\,(n-1)} 
\left[(2n +1)\,\alpha(\bq_1,\bq_2)\,F^{(n-m)}(\bk_{m+1}, ... , \bk_n)  + 2\,\beta(\bq_1,\bq_2)\,G^{(n-m)}(\bk_{m+1}, ... , \bk_n)\right]\\
 G^{(n)}(\bk_1, ... , \bk_n) &=& \sum_{m=1}^{n-1}\frac{G^{(m)}(\bk_1, ... , \bk_m)}{(2n+1)\,(n-1)} 
\left[3\,\alpha(\bq_1,\bq_2)\,F^{(n-m)}(\bk_{m+1}, ... , \bk_n)  + 2\,n\,G^{(n-m)}(\bk_{m+1}, ... , \bk_n)\right]\hspace{0.1 cm},
\ea
where $\bq_1 = \bk_1 + ... + \bk_m,\, \bq_2 = \bk_{m+1} + ... + \bk_n$ and $F^{(1)}=G^{(1)} = 1$. 
The explicit symmetrised expressions for the second order perturbation theory kernels take a very simple and intuitive form:
\ba
F^{(2)}(\bk_1,\bk_2) &=& \frac 57 + \frac 12 \frac{\bk_1\cdot \bk_2}{k_1 k_2}\left( \frac{k_1}{k_2} + \frac{k_2}{k_1}\right) 
+ \frac27 \frac{(\bk_1\cdot \bk_2)^2}{k_1^2 k_2^2}\\
G^{(2)}(\bk_1,\bk_2) &=& \frac 37 + \frac 12 \frac{\bk_1\cdot \bk_2}{k_1 k_2}\left( \frac{k_1}{k_2} + \frac{k_2}{k_1}\right) 
+ \frac47 \frac{(\bk_1\cdot \bk_2)^2}{k_1^2 k_2^2}\hspace{0.1 cm}\hspace{0.1 cm}.
\ea
One can see that mode-coupling to second order reaches its maximum when the contributing modes $\bk_1$ and  $\bk_2$ are aligned, 
whereas the kernel vanishes for anti-parallel modes. When in eqn.~(\ref{pt-modes}) $n$ different modes $\mb q_1...\mb q_n$ contribute to a mode $\mb k$, momentum 
conservation holds, enforced by the $\dirac$-distribution: $\mb k = \mb q_1+...+\mb q_n$. Similar to the renormalisation vertex in eqn.~(\ref{vertex}) also this 
motivates a diagrammatic description for the mode coupling processes (see Fig.~\ref{diagrams_1}). 

\subsection{The Four-Point Correlator in Perturbation Theory}\label{sect_pt-trispectrum}
For an analytic expression of the perturbation theory four-point correlator one has to expand the fields in the correlator. Due to the Gaussianity of the 
initial fields $\ph^{(1)}$ the correlators with an even number of fields $\ph^{(1)}$ will later simplify to products of initial power spectra 
$P_\mathrm{L}$ while all uneven contributions vanish: 
\be
\bra \ph_1\,\ph_2\,\ph_3\,\ph_4 \ket = \bra (\ph_1^{(1)}+\e^{\eta}\,\ph_1^{(2)}+...)\,(\ph_2^{(1)}+\e^{\eta}\,\ph_2^{(2)}+...)\,
(\ph_3^{(1)}+\e^{\eta}\,\ph_3^{(2)}+...)\,(\ph_4^{(1)}+\e^{\eta}\,\ph_4^{(2)}+...)\ket \hspace{0.1 cm}.
\ee
Simple truncation of the expansion in eqn.~(\ref{pt-expansion}) would lead to an inconsistent inclusion of powers of the linear power spectrum 
$P_\mathrm L^{\,k}$. We take into account all terms up to third order in the linear power spectrum which is equivalent to including terms 
with initial fields up to sixth order. Since the disconnected part will be represented by the full power spectra in eqn.~(\ref{four-point}), 
we are interested in the connected part of the correlator only. The connected part can be split into two contributions
\be
\bra \ph_1\,\ph_2\,\ph_3\,\ph_4 \ket_\mathrm c = \bra (\ph_1^{(2)}\,\ph_2^{(2)})\,(\ph_3^{(1)}\,\ph_4^{(1)}) \ket_\mathrm c + \hbox{all pairs}\in \{1,\,2,\,3,\,4\} 
			  \, + \, \bra \ph_1^{(3)}\,\ph_2^{(1)}\,\ph_3^{(1)}\,\ph_4^{(1)} \ket_\mathrm c + \mathrm{c.p.}\,\{1,\,2,\,3,\,4\}\hspace{0.1 cm}.
\ee
The first contribution originates from second order perturbation theory. In this case two of the fields in the correlator have been 
expanded to second order. The expressions in terms of the initial power spectra and the second order kernels are of the type
\be
\label{t22}
t^{\,(2)}_{a_1a_2} ((\bk_1,\bk_2),(\bk_3,\bk_4))=  4\,\e^{2\eta}\,\dirac(\bk_{1...4})\,\,P_\mathrm{L}^{\,k_3} P_\mathrm{L}^{\,k_4}\, 
 \left(\,F^{(2)}_{a_1}(\bk_{13},-\bk_3)\, F^{(2)}_{a_2}(\bk_{24},-\bk_4)\, P_\mathrm{L}^{\,k_{13}}
+ F^{(2)}_{a_1}(\bk_{14},-\bk_4)\, F^{(2)}_{a_2}(\bk_{23},-\bk_3)\, P_\mathrm{L}^{\,k_{14}}\,\right)\hspace{0.1 cm}.
\ee
The second contribution is due to third order perturbation theory. Here, one field is expanded to third order while the other three remain at linear order. 
For this reason only one perturbation kernel appears in the expression for this type of contributions:
\be
\label{t3}
t^{\,(3)}_{a_1} (\bk_1,\bk_2,\bk_3,\bk_4) = 6 \,\e^{2\eta}\, \dirac(\bk_{1...4})\,\, F^{(3)}_{a_1}(\bk_1,\bk_2,\bk_3)\, 
P_\mathrm{L}^{\,k_{1}} P_\mathrm{L}^{\,k_2} P_\mathrm{L}^{\,k_3}\hspace{0.1 cm}.
\ee
With these two functions the connected perturbation theory four-point correlator up to third order in the linear power spectrum $P_\mathrm{L}^{\,k}$ can be 
expressed by the following two tree-level contributions
\be
\label{pt-four-point}
\bra \ph_1\,\ph_2\,\ph_3\,\ph_4 \ket_\mathrm c = t^{\,(2)}_{a_1a_2} (\,(\bk_1,\,\bk_2),(\bk_3,\,\bk_4)\,) +\, \hbox{all pairs}\in \left\{1,\,2,\,3,\,4\right\} 
			\, + \, t^{\,(3)}_{a_1}(\bk_1,\,\bk_2,\,\bk_3,\,\bk_4) +\, \mathrm{c.p.}\,\{1,\,2,\,3,\,4\}\hspace{0.1 cm}.
\ee
\subsection{Trispectrum in TRG}\label{sect_trispectrum-trg}
Our main objective is to investigate the influence of the perturbation theory trispectrum on the evolution of the power spectrum. Writing the connected 
four-point correlator in terms of the trispectrum $T_{a_1 a_2 a_3 a_4}^{\bk_1, \bk_2, \bk_3}$,  
\be
\bra \ph_1\,\ph_2\,\ph_3\,\ph_4 \ket_\mathrm c = \dirac(\bk_{1...4})\,T_{a_1 a_2 a_3 a_4}^{\bk_1, \bk_2, \bk_3}\hspace{0.1 cm},
\ee
we can now include the corresponding corrections into our formalism. Taking the trispectrum in the hierarchy of eqn.~(\ref{hierarchy}) into account will 
change the closed system of eqn.~(\ref{system}) to
\ba
\label{system_tri}
\partial_\eta P_{ab}^{\,k} &=& -~\Omega_{ac}\,P_{cb}^{\,k} -\Omega_{bc}\,P_{ac}^{\,k} + \e^\eta  \int \mathrm d^3 q 
\left[\gamma_{acd}^{\,k,q,p} B_{bcd}^{\,k,q,p} +  (a\leftrightarrow b)\right]\nn \\
\partial_\eta B_{abc}^{\,k,q,p}&=&-~\Omega_{ad}B_{dbc}^{\,k,q,p}-\Omega_{bd}B_{adc}^{\,k,q,p} -\Omega_{cd}B_{abd}^{\,k,q,p} + ~2\, \e^\eta\,
\left[\gamma_{ade}^{\,k,q,p}P_{db}^{\,q}P_{ec}^{\,p}+\gamma_{bde}^{\,q,p,k}P_{dc}^{\,p}P_{ea}^{\,k}+\gamma_{cde}^{\,p,k,q}P_{da}^{\,k}P_{eb}^{\,q}\right]\nn\\
~&~& + \e^\eta\int \mathrm d^3 r \left[ 
 \gamma_{bgh}^{\,k,r,|\br-\bk|\,}T_{efgh}^{\,\bq,\bp,\br}
+\gamma_{egh}^{\,q,r,|\br-\bq|\,}T_{fbgh}^{\,\bp,\bk,\br}
+\gamma_{fgh}^{\,p,r,|\br-\bp|\,}T_{begh}^{\,\bk,\bq,\br}\right]
\hspace{0.1 cm}.
\ea
\begin{figure}
\begin{center}
\resizebox{0.7\hsize}{!}{\includegraphics{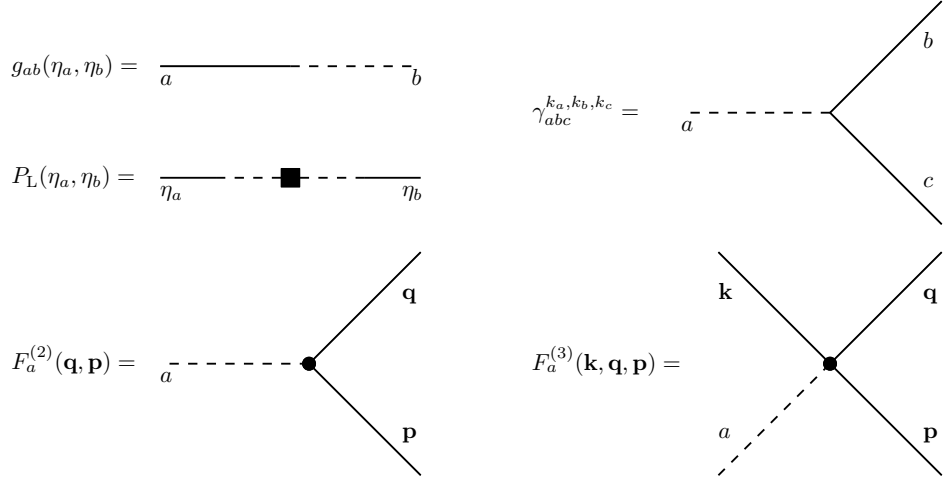}}
\caption{Diagrammatic representations of the linear propagator $g_{ab}(\eta_a,\eta_b)$, the linear power spectrum to different times 
$P_\mathrm L^{\,k}(\eta_a,\eta_b)$, the renormalisation vertex $\gamma_{abc}^{\,k_a,k_b,k_c}$ and the perturbation theory kernels 
$F_a^{(2)}(\mb q, \mb p)$ and $F_a^{(3)}(\mb k,\mb q, \mb p)$.}
\label{diagrams_1}
\end{center}
\end{figure}
Since the trispectrum is taken from perturbation theory, the evolution of the next higher correlator is not needed for its description.
One can stick to the same numerical solving procedure as it was presented in section (\ref{sect_solving}). Modifications appear in the time evolution of 
the integrals $I_{acd,bef}^{\,k}$ from eqn.~(\ref{i_evolution}) via changing the correction integrals $A_{acd,bef}^{\,k}$ from 
eqn.~(\ref{a_integral}):
\ba
A_{acd,bef}^{\,k} &\rightarrow & A_{acd,bef}^{\,k} + \Delta A_{acd,bef}^{\,k}\\
\label{a_integral_tri}
\Delta A_{acd,bef}^{\,k} &=&  \frac {k}{8\,\pi} \int \mathrm d^3 q\int \mathrm d^3 r
\left \{ \,
\gamma_{acd}^{k,q,p} \left [ 
 \gamma_{bgh}^{\,k,r,|\br-\bk|\,}T_{efgh}^{\,\bq,\bp,\br}
+\gamma_{egh}^{\,q,r,|\br-\bq|\,}T_{fbgh}^{\,\bp,\bk,\br}
+\gamma_{fgh}^{\,p,r,|\br-\bp|\,}T_{begh}^{\,\bk,\bq,\br} + \left(q \leftrightarrow p\right)
			  \right ]
\,\right \} \hspace{0.1 cm}.
\ea
While the former expression in eqn.~(\ref{a_integral}) was a one loop integral only, we now have to integrate twice over the full $k$-space. The reason for this are 
the additional $\dirac$-functions in the disonnected parts of the four-point correlator in eqn.~(\ref{four-point}). 
The integration is performed numerically using Monte Carlo integration techniques from the multidimensional numerical integration library CUBA \citep{Hahn2005}.
\section{Diagrammatic Description}\label{sect_diagrams}
An analytic solution for the system in eqn.~(\ref{system}) can be formulated \citep{Pietroni2008}. This is still the case for the system in eqn.~(\ref{system_tri}) 
with additional trispectrum terms. Solving first the linearised evolution equations~(\ref{sfe_compact}) one can write down the linear solutions for the fields $\ph_{a,\,\mathrm L}(\bk,\eta)$ 
with the help of the linear propagator $g_{ab}(\eta,\eta^\prime)$ \citep{Matarrese2007, Crocce2006a} :
\be
 \ph_{a,\,\mathrm L}(\bk,\eta) = g_{ab}(\eta,\eta^\prime)\,\ph_{b,\,\mathrm L}(\bk, \eta^\prime)\hspace{0.1 cm}.
\ee
Furthermore, the linear propagator has the following properties:
\ba
 \partial_\eta\,g_{ab}(\eta,\eta^\prime) &=& -\Omega_{ac}(\eta)\,g_{cb}(\eta,\eta^\prime)\nn\\
 g_{ab}(\eta,\eta^\prime) &=& \delta_{ab}\nn\\
 g_{ab}(\eta,\eta^\prime)\, g_{bc}(\eta^\prime,\eta^{\prime\prime}) &=& g_{ac}(\eta,\eta^{\prime\prime})\hspace{0.1cm}.
\ea
With help of this linear propagator a formal analytic solution can be given for the system~(\ref{system_tri}):
\ba
\label{solution_p}
P_{ab}^{\,k}(\eta)&=&g_{ac}(\eta,0)\,g_{bd}(\eta,0)\,\, P_{cd}^{\,k}(\eta=0) + \int_0^{\,\eta}\, \mathrm d \eta^\prime \e^{\eta^\prime}\int \mathrm d ^3 q 
\,\,g_{ae}(\eta,\eta^\prime)\,g_{bf}(\eta,\eta^\prime) 
\left[\gamma_{ecd}^{\,k,q,p}B_{fcd}^{\,k,q,p}(\eta^\prime)+\gamma_{fcd}^{\,k,q,p}B_{ecd}^{\,k,q,p}(\eta^\prime)\right]\\
B_{abc}^{\,k,q,p}(\eta)&=&g_{ad}(\eta,0)\,g_{be}(\eta,0)\,g_{cf}(\eta,0)\,\, B_{abc}^{\,k,q,p}(\eta=0)\nn\\
 ~&~&+\int_0^{\,\eta}\, \mathrm d \eta^\prime \e^{\eta^\prime}\int \mathrm d ^3 q \,\, \left[ \quad 2\,\left(
\gamma_{dgh}^{\,k,q,p}\,P_{eg}^{\,q}(\eta^\prime)\,P_{fh}^{\,p}(\eta^\prime) +
\gamma_{egh}^{\,q,p,k}\,P_{fg}^{\,p}(\eta^\prime)\,P_{dh}^{\,k}(\eta^\prime) +
\gamma_{fgh}^{\,p,k,q}\,P_{dg}^{\,k}(\eta^\prime)\,P_{eh}^{\,q}(\eta^\prime) \right) \right.\nn\\
~&~&\hspace{3cm}+\int\mathrm d^3 r \left. \left( 
\gamma_{dgh}^{\,k,r,|\br-\bk|\,}T_{efgh}^{\,\bq,\bp,\br}
+\gamma_{egh}^{\,q,r,|\br-\bq|\,}T_{fdgh}^{\,\bp,\bk,\br}
+\gamma_{fgh}^{\,p,r,|\br-\bp|\,}T_{degh}^{\,\bk,\bq,\br} \right) \quad \right]\hspace{0.1 cm}.
\label{solution_b}
\ea
\begin{figure}
\begin{center}
\resizebox{0.6\hsize}{!}{\includegraphics{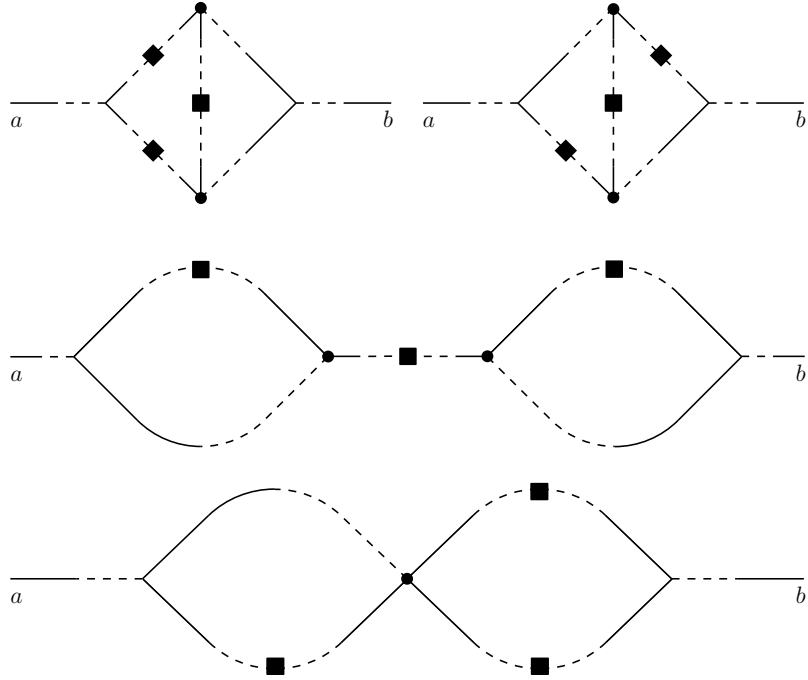}}
\caption{Two loop corrections to the power spectrum originating from the trispectrum of tree-level perturbation theory in eqn.~(\ref{pt-four-point}). 
The first three diagrams are due to second order perturbations from eqn.~(\ref{t22}). Third order perturbation theory in eqn.~(\ref{t3}) leads to the last diagram. 
While the perturbation theory was included on tree-level only, the non-perturbative time renormalisation leads to 2-loop corrections to the power spectrum. 
All diagrams are third order in the initial power spectrum and second order in the renormalisation vertex $\mathcal O(\gamma)$. The latter is due to the fact that 
both evolution equations in (\ref{hierarchy}) are first order in the vertex.}
\label{diagrams_2}
\end{center}
\end{figure}
For a better understanding of the tripectrum corrections to the power spectrum it is useful to analyse the equations in a 
diagrammatic representation. In the left panel of Fig.~\ref{diagrams_1} symbols for the linear propagator $g_{ab}(\eta_a,\eta_b)$, the renormalization vertex 
$\gamma_{abc}^{\,k_a,k_b,k_3}$ and the linear power spectrum to different times $P_\mathrm L (\eta_a,\eta_b)$ are depicted. A diagrammatic representation for the 
perturbation theory kernels are also needed to describe the trispectrum. Since we used the perturbation theory trispectrum to third order in the power spectrum, 
the kernel of second and third order are sufficient (Fig.~\ref{diagrams_1}). As one can see from eqs.~(\ref{vertex}) and (\ref{pt-modes}) 
momentum conservation holds at each intersection in the diagrams. 

The perturbation theory trispectrum terms lead to corrections in the bispectrum in eqn.~({\ref{solution_b}}). In eqn.~(\ref{solution_p}) these terms are transported to 
corrections to the power spectrum itself. The types of corrections to the power spectrum are shown in Fig.~\ref{diagrams_2}. All the terms are second order 
corrections in the renormalisation vertex $\mathcal O (\gamma^2)$ since both evolution equations are first order in $\gamma$. The first three diagrams in 
Fig.~\ref{diagrams_2} originate from the second order kernels $F_a^{(2)}$ in eqn.~(\ref{t22}), while the last diagramm is due to the third order perturbation kernel 
$F_a^{(3)}$ in eqn.~(\ref{t3}). Also here, one can see that all perturbative corrections we included are of third order in the linear power spectrum. 
While the perturbation theory trispectrum is calculated at tree-level, both the evolution equations - for the bispectrum and for the power spectrum - introduce 
one momentum integration. Therefore,  all these correction diagrams are two loop diagrams. 

We want to emphazise at this point that the inclusion of the perturbation theory trispectrum does not lead to a simple perturbative correction only. 
At each time step the perturbative trispectrum corrects the evolution of bispectrum and power spectrum. 
Therefore, from that moment on these corrections will be involved in the non-perturbative method of time renormalisation. 
In this work we only discuss the trispectrum corrections to this method, since the quality and performance of the original time renormalisation 
technique has been thoroughly discussed already \citep{Pietroni2008}. 
\begin{figure}
\begin{center}
 \resizebox{0.49\hsize}{!}{\includegraphics{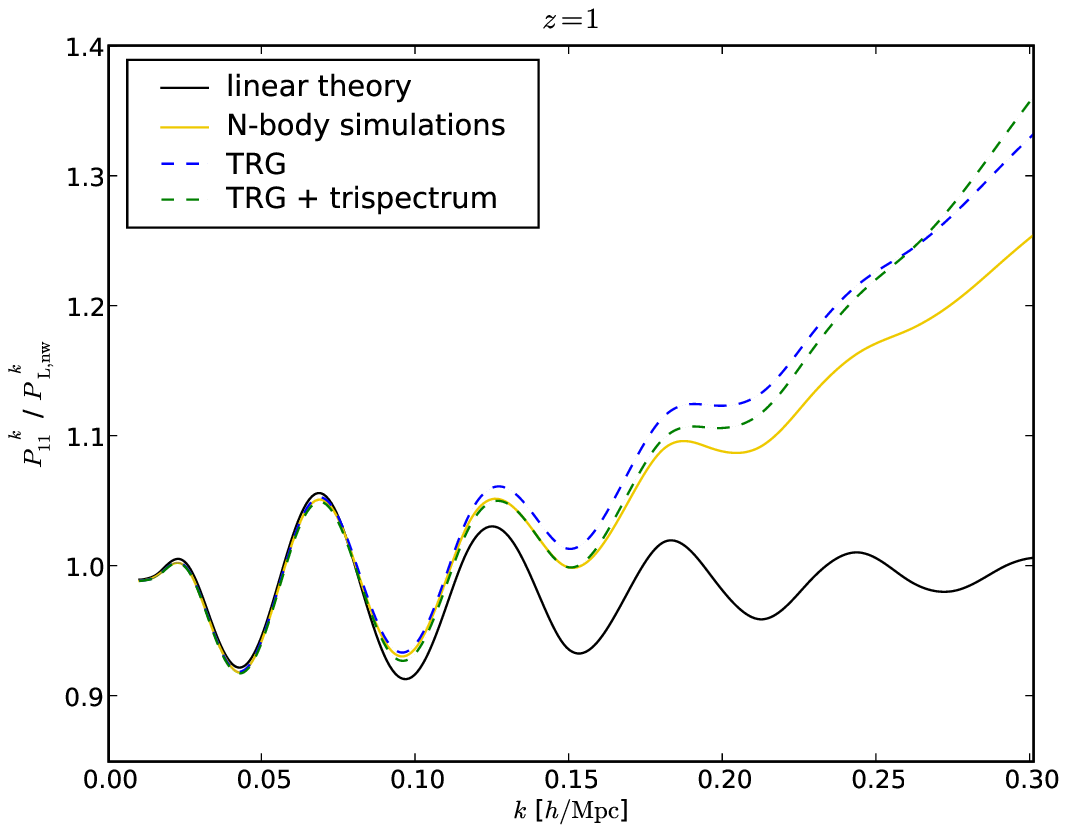}}\quad
 \resizebox{0.49\hsize}{!}{\includegraphics{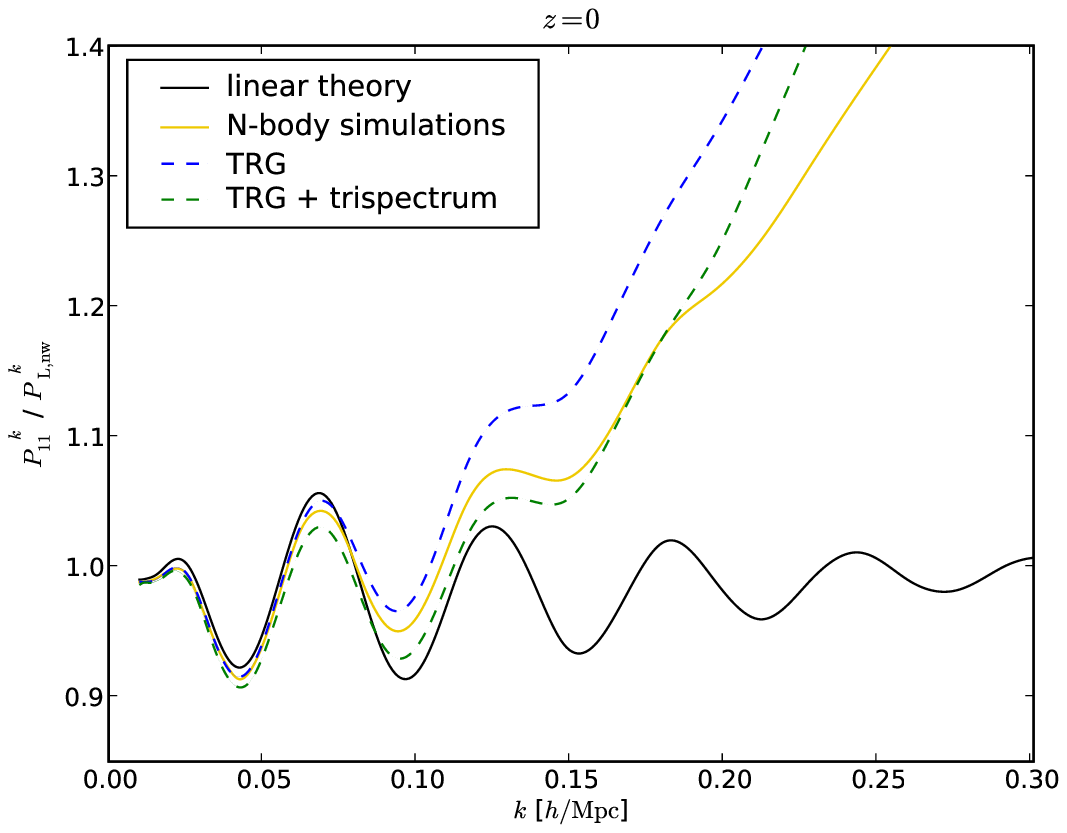}}
\caption{Matter power spectra $P_{11}^{\,k}$ divided by a linear spectrum without baryonic wiggles $P_\mathrm{L,nw}^{\,k}$ \citep{Eisenstein1998a} 
in the mildly non-linear regime. 
The linear spetrum is depicted as solid (black) line. A spectrum from $N$-body simulations \citep{Carlson2009} serves as reference (yellow). The dashed 
(blue) line is time-RG theory and the dashed (green) line is time-RG theory with trispectrum included. Left panel: $z=1$, right panel: $z=0$.}
\label{fig1}
\end{center}
\end{figure}
\section{Numerical Results}\label{sect_results}
We solved the system from equation~(\ref{system_tri}) starting from redshift $z=100$ well within the linear regime with linear initial power spectra 
(see eqn.~(\ref{plin})) and vanishing bispectrum. We evolved the system to redshift $z=1$ and $z=0$ with and without trispectrum included and 
compared the power spectra with results from numerical simulations of the same $\Lambda$CDM cosmology \citep{Carlson2009}. \gerot{Due to numerical complexity only 
power spectra up to $k=1 \,h\mathrm{Mpc}^{-1}$ were included in the trispectrum integrals $\Delta I_{acd,bef}^{\,k}$ from eqn.~(\ref{a_integral_tri}). However, in 
the integrals $I_{acd,bef}^{\,k}$ from eqn.~(\ref{a_integral}) modes up to $k=10 \,h\mathrm{Mpc}^{-1}$ were taken into account, were the results saturate to 
percent accuracy.}

The results are shown in Fig.~\ref{fig1}, in which also the linearly evolved power spectrum is depicted. 
All spectra were divided by a linear power spectrum $P_\mathrm{L,nw}^{\,k}$ without wiggles from baryonic acoustic oscillations 
\citep[][eqn.~29]{Eisenstein1998a}. 
For $z=1$ the results with the trispectrum included are in excellent agreement with numerical simulations up to $k\simeq0.17 
\,h\mathrm{Mpc}^{-1}$. For $0.17\,h\mathrm{Mpc}^{-1} \leq k\leq0.25 \,h\mathrm{Mpc}^{-1}$ the method performs still better than without trispectrum included. 
Beyond this regime the perturbative description of the trispectrum breaks down and the results are in strong disagreement with simulations. 

It is not surprising that below a certain scale the method performs better without the perturbative trispectrum included. For larger $k$ in the 
integrals $\Delta A_{acd,bef}^{\,k}$ in eqn.~(\ref{a_integral_tri}) also trispectra at smaller scales contribute to the evolution. 
Since in perturbation theory loop corrections become more and more important at smaller scales, the tree-level trispectrum description breaks down above a specific 
wave number. For this reason, beyond $ k\simeq 0.25 \,h\mathrm{Mpc}^{-1}$ time renormalisation without trispectrum will lead to better results in comparisons 
to numerical simulations.

\gerot{At $z=0$ trispectrum corrections overcompensate the too strong growth in the pure TRG approach on scales below $k \approx 0.15 \,h\mathrm{Mpc}^{-1}$ and lead to 
too little growth. In this regime our results agree with the numerical simulations within $2.5\%$. The better agreement for $z=1$ on these scales may simply be due to the 
breakdown of the tree-level perturbative descrition of the trispectrum at later times.
Beyond $k \approx 0.15 \,h\mathrm{Mpc}^{-1}$ the inclusion of the trispectrum leads to a better agreement with simulations than pure TRG, 
while both methods show too strong growth further inside the non-linear regime. }

\gerot{The results for pure TRG may differ from the results of \cite{Pietroni2008}, since only 12 instead of 
14 independent integrals $I_{acd,bef}^{\,k}$ were included in the original method. However, in later applications all 14 independent were taken into account.}

\del{While the perturbation theory trispectrum will not lead to a step forward beyond $ k\simeq 0.25 \,h\mathrm{Mpc}^{-1}$, the method performs remarkably well 
on larger scales at $z=1$. At $z=0$ the agreement is not as remarkable as at $z=1$ but the deviation from 
the reference power spectrum from numerical simulations stays within $2.5\%$.}

\section{Summary}\label{sect_summary}
In this work the influence of the tree-level trispectrum on the time renormalisation method \citep{Pietroni2008} has been studied. 
To keep the computational effort at a reasonable level we did not include the full trispectrum evolution from the hierarchy. 
Instead, we used tree-level perturbation theory for the trispectrum in the evolution of the bispectrum.

(1) The trispetrum was formulated in standard perturbation theory to third order in the linear power spectrum $P_\mathrm L ^{\,k}$. 
This was included in the evolution equation of the bispectrum. 

(2) The numerical method for solving the system was extended by the trispectrum corrections. Due to the linear time dependence of the perturbation theory 
trispectrum it is sufficient to calculate the correction integals at one fixed initial time. Once the corrections are derived the extended method operates at the same 
speed as the time renormalisation method without trispectrum included. The additional 2-loop corrections to the power spectrum are of second order in the renormalisation 
vertex $\mathcal O(\gamma^2)$ and third order in the linear power spectrum $\mathcal O(P_\mathrm L ^3)$.

(3) Perturbative trispectrum corrections are fed into the systems at all times. Once a correction has been included its evolution will be described 
by the non-perturbative formalism of time renormalisation. Therefore, although the trispectrum was only taken into account perturbatively, 
its inclusion can not be interpreted as a pure perturbative correction disentangled from renormalisation. 

(4) \gerot{We solved the system numerically starting with Gaussian initial conditions and a linear power spectrum at an initial redshift of $z=100$. In comparison 
to numerical simulations \citep{Carlson2009} the inclusions of the trispectrum generally improves the results up to $k\approx 0.25\,h\mathrm{Mpc}^{-1}$. 
However, on larger scales the damping due to perturbative tree-level trispectrum overcompensates the deviation of pure TRG from N-body simulations.
The results agree with the simulations within 1\% up to $k\approx 0.18\,h\mathrm{Mpc}^{-1}$ for $z=1$ and within 2.5\% up to $k\approx 0.2\,h\mathrm{Mpc}^{-1}$ for $z=0$.}

(5) Beyond $k\approx 0.25\,h\mathrm{Mpc}^{-1}$ the perturbative description of the trispectrum breaks down and the method performs better without trispectrum included. 
This is due to the fact that loop corrections to the trispectrum are not included in our method and become more and more important on smaller scales. 
Adding the perturbation theory trispectrum therefore predominantly pays off on large scales and at the beginning of the mildly non-linear regime.

(6) Although the prediction of the amplitude and position of the first two peaks in the baryonic acoustic oscillations was improved by the trispectrum, we are far from 
reaching 1\% accuracy over the entire BAO regime. Also the speed of this method was reduced by including the trispectrum, since at a specific time the 2-loop trispectrum 
corrections have to be derived. Finding a better analytical estimator for the trispectrum from other renormalisation approaches, which includes higher 
order corrections, could improve the results of this method further inside the mildly non-linear regime. 
 
\section*{Acknowledgements}
We would like to thank Bj\"orn Malte Sch\"afer for providing the cosmological code basement and helpful advice in many difficult questions. 
We also thank Massimo Pietroni for useful discussions and for providing reference data of power spectra. 
Our work was supported by the German Research Foundation (DFG) within the framework of the Priority Programme 1177.

\bibliography{bibtex/aamnem,bibtex/references}

\begin{thebibliography}{}

\bibitem[\protect\citeauthoryear{{Angulo}, {Baugh}, {Frenk} \&
  {Lacey}}{{Angulo} et~al.}{2008}]{Angulo2008}
{Angulo} R.~E.,  {Baugh} C.~M.,  {Frenk} C.~S.,    {Lacey} C.~G.,  2008,
  \mnras, 383, 755

\bibitem[\protect\citeauthoryear{{Anselmi}, {Matarrese} \&
  {Pietroni}}{{Anselmi} et~al.}{2011}]{Anselmi2011}
{Anselmi} S.,  {Matarrese} S.,    {Pietroni} M.,  2011, \jcap, 6, 15

\bibitem[\protect\citeauthoryear{{Bernardeau}, {Colombi}, {Gazta{\~n}aga} \&
  {Scoccimarro}}{{Bernardeau} et~al.}{2002}]{Bernardeau2002}
{Bernardeau} F.,  {Colombi} S.,  {Gazta{\~n}aga} E.,    {Scoccimarro} R.,
  2002, \physrep, 367, 1

\bibitem[\protect\citeauthoryear{{Bernardeau}, {Crocce} \&
  {Scoccimarro}}{{Bernardeau} et~al.}{2008}]{Bernardeau2008}
{Bernardeau} F.,  {Crocce} M.,    {Scoccimarro} R.,  2008, \prd, 78, 103521

\bibitem[\protect\citeauthoryear{{Blake}, {Collister}, {Bridle} \&
  {Lahav}}{{Blake} et~al.}{2007}]{Blake2007}
{Blake} C.,  {Collister} A.,  {Bridle} S.,    {Lahav} O.,  2007, \mnras, 374,
  1527

\bibitem[\protect\citeauthoryear{{Carlson}, {White} \& {Padmanabhan}}{{Carlson}
  et~al.}{2009}]{Carlson2009}
{Carlson} J.,  {White} M.,    {Padmanabhan} N.,  2009, \prd, 80, 043531

\bibitem[\protect\citeauthoryear{{Crocce} \& {Scoccimarro}}{{Crocce} \&
  {Scoccimarro}}{2006a}]{Crocce2006}
{Crocce} M.,  {Scoccimarro} R.,  2006a, \prd, 73, 063520

\bibitem[\protect\citeauthoryear{{Crocce} \& {Scoccimarro}}{{Crocce} \&
  {Scoccimarro}}{2006b}]{Crocce2006a}
{Crocce} M.,  {Scoccimarro} R.,  2006b, \prd, 73, 063519

\bibitem[\protect\citeauthoryear{{Crocce} \& {Scoccimarro}}{{Crocce} \&
  {Scoccimarro}}{2008}]{Crocce2008}
{Crocce} M.,  {Scoccimarro} R.,  2008, \prd, 77, 023533

\bibitem[\protect\citeauthoryear{{Eisenstein} \& {Hu}}{{Eisenstein} \&
  {Hu}}{1998}]{Eisenstein1998a}
{Eisenstein} D.~J.,  {Hu} W.,  1998, \apj, 496, 605

\bibitem[\protect\citeauthoryear{{Eisenstein}, {Hu} \& {Tegmark}}{{Eisenstein}
  et~al.}{1998}]{Eisenstein1998}
{Eisenstein} D.~J.,  {Hu} W.,    {Tegmark} M.,  1998, \apjl, 504, L57

\bibitem[\protect\citeauthoryear{{Eisenstein}, {Seo} \& {White}}{{Eisenstein}
  et~al.}{2007}]{Eisenstein2007}
{Eisenstein} D.~J.,  {Seo} H.-J.,    {White} M.,  2007, \apj, 664, 660

\bibitem[\protect\citeauthoryear{{Eisenstein}, {Zehavi}, {Hogg}, {Scoccimarro},
  {Blanton}, {Nichol} \& {Scranton}}{{Eisenstein}
  et~al.}{2005}]{Eisenstein2005}
{Eisenstein} D.~J.,  {Zehavi} I.,  {Hogg} D.~W.,  {Scoccimarro} R.,  {Blanton}
  M.~R.,  {Nichol} R.~C.,    {Scranton} 2005, \apj, 633, 560

\bibitem[\protect\citeauthoryear{{Evrard}, {Bialek}, {Busha}, {White}, {Habib},
  {Heitmann}, {Warren}, {Rasia}, {Tormen}, {Moscardini}, {Power}, {Jenkins},
  {Gao}, {Frenk}, {Springel}, {White} \& {Diemand}}{{Evrard}
  et~al.}{2008}]{Evrard2008}
{Evrard} A.~E.,  {Bialek} J.,  {Busha} M.,  {White} M.,  {Habib} S.,
  {Heitmann} K.,  {Warren} M.,  {Rasia} E.,  {Tormen} G.,  {Moscardini} L.,
  {Power} C.,  {Jenkins} A.~R.,  {Gao} L.,  {Frenk} C.~S.,  {Springel} V.,
  {White} S.~D.~M.,    {Diemand} J.,  2008, \apj, 672, 122

\bibitem[\protect\citeauthoryear{{Glazebrook}, {Eisenstein}, {Dey}, {Nichol} \&
  {The WFMOS Feasibility Study Dark Energy Team}}{{Glazebrook}
  et~al.}{2005}]{Glazebrook2005}
{Glazebrook} K.,  {Eisenstein} D.,  {Dey} A.,  {Nichol} B.,    {The WFMOS
  Feasibility Study Dark Energy Team} 2005, ArXiv Astrophysics e-prints

\bibitem[\protect\citeauthoryear{{Goroff}, {Grinstein}, {Rey} \&
  {Wise}}{{Goroff} et~al.}{1986}]{Goroff1986}
{Goroff} M.~H.,  {Grinstein} B.,  {Rey} S.-J.,    {Wise} M.~B.,  1986, \apj,
  311, 6

\bibitem[\protect\citeauthoryear{{Guo} \& {Jing}}{{Guo} \&
  {Jing}}{2009}]{Guo2009}
{Guo} H.,  {Jing} Y.~P.,  2009, \apj, 698, 479

\bibitem[\protect\citeauthoryear{{Hahn}}{{Hahn}}{2005}]{Hahn2005}
{Hahn} T.,  2005, Computer Physics Communications, 168, 78

\bibitem[\protect\citeauthoryear{{Heitmann}, {Luki{\'c}}, {Fasel}, {Habib},
  {Warren}, {White}, {Ahrens}, {Ankeny}, {Armstrong}, {O'Shea}, {Ricker},
  {Springel}, {Stadel} \& {Trac}}{{Heitmann} et~al.}{2008}]{Heitmann2008}
{Heitmann} K.,  {Luki{\'c}} Z.,  {Fasel} P.,  {Habib} S.,  {Warren} M.~S.,
  {White} M.,  {Ahrens} J.,  {Ankeny} L.,  {Armstrong} R.,  {O'Shea} B.,
  {Ricker} P.~M.,  {Springel} V.,  {Stadel} J.,    {Trac} H.,  2008,
  Computational Science and Discovery, 1, 015003

\bibitem[\protect\citeauthoryear{{Heitmann}, {White}, {Wagner}, {Habib} \&
  {Higdon}}{{Heitmann} et~al.}{2010}]{Heitmann2010}
{Heitmann} K.,  {White} M.,  {Wagner} C.,  {Habib} S.,    {Higdon} D.,  2010,
  \apj, 715, 104

\bibitem[\protect\citeauthoryear{{Hill}, {Gebhardt}, {Komatsu}, {Drory},
  {MacQueen}, {Adams}, {Blanc}, {Koehler}, {Rafal}, {Roth}, {Kelz}, {Gronwall},
  {Ciardullo} \& {Schneider}}{{Hill} et~al.}{2008}]{Hill2008}
{Hill} G.~J.,  {Gebhardt} K.,  {Komatsu} E.,  {Drory} N.,  {MacQueen} P.~J.,
  {Adams} J.,  {Blanc} G.~A.,  {Koehler} R.,  {Rafal} M.,  {Roth} M.~M.,
  {Kelz} A.,  {Gronwall} C.,  {Ciardullo} R.,    {Schneider} D.~P.,  2008, 399,
  115

\bibitem[\protect\citeauthoryear{{Huff}, {Schulz}, {White}, {Schlegel} \&
  {Warren}}{{Huff} et~al.}{2007}]{Huff2007}
{Huff} E.,  {Schulz} A.~E.,  {White} M.,  {Schlegel} D.~J.,    {Warren} M.~S.,
  2007, Astroparticle Physics, 26, 351

\bibitem[\protect\citeauthoryear{{H{\"u}tsi}}{{H{\"u}tsi}}{2006}]{Hutsi2006}
{H{\"u}tsi} G.,  2006, \aap, 449, 891

\bibitem[\protect\citeauthoryear{{Jain} \& {Bertschinger}}{{Jain} \&
  {Bertschinger}}{1994}]{Jain1994}
{Jain} B.,  {Bertschinger} E.,  1994, \apj, 431, 495

\bibitem[\protect\citeauthoryear{{Jeong} \& {Komatsu}}{{Jeong} \&
  {Komatsu}}{2006}]{Jeong2006}
{Jeong} D.,  {Komatsu} E.,  2006, \apj, 651, 619

\bibitem[\protect\citeauthoryear{{Jeong} \& {Komatsu}}{{Jeong} \&
  {Komatsu}}{2009}]{Jeong2009}
{Jeong} D.,  {Komatsu} E.,  2009, \apj, 691, 569

\bibitem[\protect\citeauthoryear{{Lesgourgues}, {Matarrese}, {Pietroni} \&
  {Riotto}}{{Lesgourgues} et~al.}{2009}]{Lesgourgues2009}
{Lesgourgues} J.,  {Matarrese} S.,  {Pietroni} M.,    {Riotto} A.,  2009,
  \jcap, 6, 17

\bibitem[\protect\citeauthoryear{{Lesgourgues} \& {Pastor}}{{Lesgourgues} \&
  {Pastor}}{2006}]{Lesgourgues2006}
{Lesgourgues} J.,  {Pastor} S.,  2006, \physrep, 429, 307

\bibitem[\protect\citeauthoryear{{Matarrese} \& {Pietroni}}{{Matarrese} \&
  {Pietroni}}{2007}]{Matarrese2007}
{Matarrese} S.,  {Pietroni} M.,  2007, \jcap, 6, 26

\bibitem[\protect\citeauthoryear{{Matarrese} \& {Pietroni}}{{Matarrese} \&
  {Pietroni}}{2008}]{Matarrese2008}
{Matarrese} S.,  {Pietroni} M.,  2008, Modern Physics Letters A, 23, 25

\bibitem[\protect\citeauthoryear{{Padmanabhan}, {Schlegel}, {Seljak},
  {Makarov}, {Bahcall} \& {Blanton}}{{Padmanabhan}
  et~al.}{2007}]{Padmanabhan2007}
{Padmanabhan} N.,  {Schlegel} D.~J.,  {Seljak} U.,  {Makarov} A.,  {Bahcall}
  N.~A.,    {Blanton} 2007, \mnras, 378, 852

\bibitem[\protect\citeauthoryear{{Peacock} \& {Dodds}}{{Peacock} \&
  {Dodds}}{1996}]{Peacock1996}
{Peacock} J.~A.,  {Dodds} S.~J.,  1996, \mnras, 280, L19

\bibitem[\protect\citeauthoryear{{Peebles}}{{Peebles}}{1980}]{Peebles1980}
{Peebles} P.~J.~E.,  1980

\bibitem[\protect\citeauthoryear{{Pietroni}}{{Pietroni}}{2008}]{Pietroni2008}
{Pietroni} M.,  2008, \jcap, 10, 36

\bibitem[\protect\citeauthoryear{{Seo} \& {Eisenstein}}{{Seo} \&
  {Eisenstein}}{2003}]{Seo2003}
{Seo} H.-J.,  {Eisenstein} D.~J.,  2003, \apj, 598, 720

\bibitem[\protect\citeauthoryear{{Smith}, {Peacock}, {Jenkins}, {White},
  {Frenk}, {Pearce}, {Thomas}, {Efstathiou} \& {Couchman}}{{Smith}
  et~al.}{2003}]{Smith2003}
{Smith} R.~E.,  {Peacock} J.~A.,  {Jenkins} A.,  {White} S.~D.~M.,  {Frenk}
  C.~S.,  {Pearce} F.~R.,  {Thomas} P.~A.,  {Efstathiou} G.,    {Couchman}
  H.~M.~P.,  2003, \mnras, 341, 1311

\bibitem[\protect\citeauthoryear{{Springel}}{{Springel}}{2005}]{Springel2005}
{Springel} V.,  2005, \mnras, 364, 1105

\bibitem[\protect\citeauthoryear{{Takahashi}, {Yoshida}, {Matsubara},
  {Sugiyama}, {Kayo}, {Nishimichi}, {Shirata}, {Taruya}, {Saito}, {Yahata} \&
  {Suto}}{{Takahashi} et~al.}{2008}]{Takahashi2008}
{Takahashi} R.,  {Yoshida} N.,  {Matsubara} T.,  {Sugiyama} N.,  {Kayo} I.,
  {Nishimichi} T.,  {Shirata} A.,  {Taruya} A.,  {Saito} S.,  {Yahata} K.,
  {Suto} Y.,  2008, \mnras, 389, 1675

\bibitem[\protect\citeauthoryear{{Tassev} \& {Zaldarriaga}}{{Tassev} \&
  {Zaldarriaga}}{2011}]{Tassev2011}
{Tassev} S.,  {Zaldarriaga} M.,  2011, ArXiv e-prints

\bibitem[\protect\citeauthoryear{{Valageas}}{{Valageas}}{2008}]{Valageas2008}
{Valageas} P.,  2008, \aap, 484, 79

\end{thebibliography}
\bibliographystyle{mn2e}

\appendix

\bsp

\label{lastpage}

\end{document}